\documentclass[superscriptaddress,
twocolumn,showpacs,preprintnumbers,amsmath,amssymb]{revtex4}
\usepackage{graphicx}
\usepackage{dcolumn}
\usepackage{bm}
\usepackage{latexsym}
\usepackage{psfrag}
\usepackage{epsfig}
\begin{document}
\title{Searching for targets on a model DNA: \\
Effects of inter-segment hopping, detachment and re-attachment
}
\author{Debanjan Chowdhury{\footnote{debanjan@iitk.ac.in}}}
\address{Department of Physics, Indian Institute of Technology, Kanpur 208016, India.}
\date{\today}%
\begin{abstract}
For most of the important processes in DNA metabolism, a protein has to reach a specific
binding site on the DNA. The specific binding site may consist of just a few base pairs
while the DNA is usually several millions of base pairs long. How does the protein search
for the target site? What is the most efficient mechanism for a successful search? Motivated
by these fundamental questions on intracellular biological processes, we have developed a
model for searching a specific site on a model DNA by a single protein. We
have made a comparative quantitative study of the efficiencies of sliding, inter-segmental
hoppings and detachment/re-attachments of the particle during its search for the specific
site on the DNA. We also introduce some new quantitative measures of {\it efficiency} of a search
process by defining a relevant quantity, which can be measured in {\it in-vitro} experiments.
\end{abstract}

\maketitle
\section{Introduction}

Self-avoiding walk (SAW) on a lattice serves as a paradigm for research 
in statistical properties of natural and artificial polymers 
\cite{degennes}. 
A ``bridge'' is defined as a bond on the lattice that connects two sites 
both of which are located on the SAW and are nearest-neighbours on the 
lattice but are not nearest neighbours along the contour of the SAW. 
RWs on SAWs is an interesting problem in its own right because of the 
interesting effects of the hops of the random walker across the bridges. 
Many years ago, motivated by the vibrational dynamics of proteins, the 
root-mean-square displacement of the random walker on a SAW was studied 
both in the absence and presence of hops across bridges 
\cite{chowdhury85a,chowdhury85b,yang85,maritan85,bouchaud86,seno89,manna89}. 
RW on SAW has also been studied as one of the prototypes of RW in 
disordered and fractal media \cite{Bouchaud,havlin,klafter07}. 

In this paper we report the effects of the hops of the random walker 
across the bridges on the distributions of their first passage times,  
(FPT) \cite{redner}, i.e., the time taken by the walker to reach a 
target site for the first time.
Moreover, we extend the model even further by allowing the possibility 
of detachments and re-attachments (to be described in detail in section 
\ref{model}); we also report the effects of these processes of 
attachments/detachments of the random walkers on the distributions of 
their first passage times. This extension of the model and the 
computation of the first passage times are motivated by a biological 
process which is discussed in the next section. Therefore, 
this work may also be viewed as a biologically motivated extension of 
the works reported earlier.\cite{chowdhury85a,chowdhury85b,yang85,maritan85,bouchaud86,seno89,manna89}.

This paper is organised as follows: In section {\ref{motivation}}, we discuss
the biological motivation behind this problem. In section {\ref{review}}, we
review some of the earlier works and compare our model to the previous models.
In section {\ref{model}}, we build the model and, thereafter in section
 {\ref{result}}, we discuss the results.  
\section{Biological motivation}
\label{motivation}

A cell is the structural and functional unit of a living system. DNA, the 
device used by nature for storage of genetic information, is essentially 
a linear polymer. The genetic information is chemically encoded in the 
sequence of the nucleotides, the monomeric subunits of DNA. Some viruses 
use RNA, instead of DNA, for storage of genetic information. In almost 
all processes involved in the nucleic acid (DNA or RNA) metabolism, 
specific proteins (or, more generally, macromolecular complexes) need to 
bind to specific sites on the nucleic acid. For example, a transcription 
factor must bind at the appropriate site on the DNA to initiate the 
process of transcription whereby genetic code is transcribed from the 
DNA to the corresponding RNA. Similarly, the processes of 
DNA replication, repair and recombination also require binding of the 
corresponding appropriate proteins at specific sites on the DNA. Other 
processes of similar nature include restriction and modification of DNA 
by sequence-specific endonucleases. The typical length of a DNA chain 
could be millions of base-pairs, whereas the target site may be a 
sequence of just a few basepairs. But, a protein usually succeeds in 
reaching the target in an unbelievably short time. One of the most 
challenging open questions in molecular cell biology and biophysics is: 
how does a protein search such a long strand of DNA in an efficient 
manner to reach the target site?

\begin{figure}[ph]
\centerline{\psfig{file=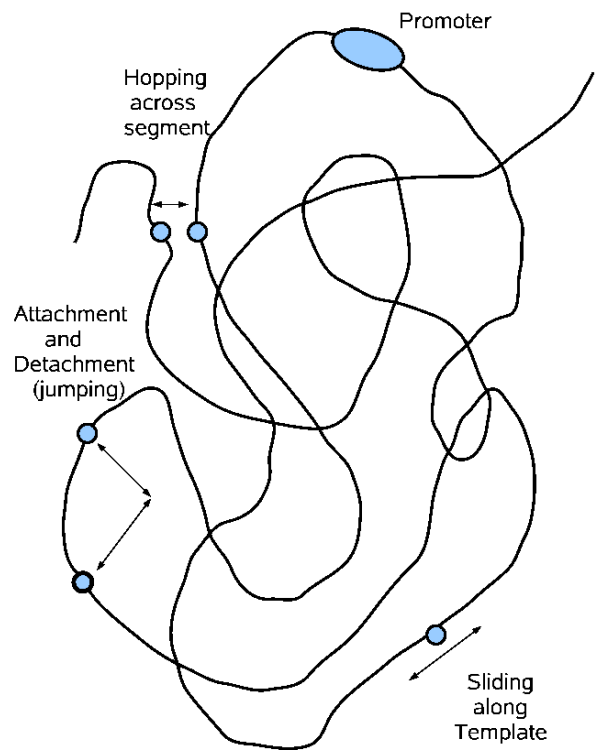,width=7.7cm}}
\vspace*{8pt}
\caption{A pictorial depiction of the various mechanisms of searching for specific binding sites by a DNA-binding 
protein (e.g., searching of the promoter site by a transcription factor).\label{rnapcartoon}}
\end{figure}

To our knowledge, this question was first formulated clearly by Von Hippel 
and coworkers \cite{Hippel1,Hippel2} who also pointed out three possible 
mechanisms of search for the specific binding sites by the DNA-binding 
proteins. These three possible modes of search are as follows: \\ 
(i) The protein {\it slides } diffusively along an effectively 
one-dimensional track formed by covalently-bonded bases of the DNA template,\\ 
(ii) it not only slides along the DNA chain but, occasionally, also 
{\it hops } from one segment of the DNA to a neighbouring segment; 
proteins with more than one DNA-binding sites can exploit this mechanism,\\ 
(iii) in addition to sliding and intersegmental hopping, it also carries 
out a three-dimensional search for the specific binding site by first 
{\it detaching} from the DNA strand and, then, after executing 
three-dimensional diffusion in the solution, {\it re-attaching} at a new 
site which is uncorrelated with the site from which it detached (see 
Fig.{\ref{rnapcartoon}}). \\
Various aspects of these mechanisms and their relative importance have 
been explored by many research groups in subsequent works (see next
section for a brief review and comparison to our model).
\cite{elf07,Busta,Halford1,Holyst,Halford2,Moreau,slutsky04,Mirny,oshanin,salerno,Zhou,Kampmann,metzler05,klafter06,lindenberg07,lindenberg08,Flyvbjerg,murugan07,sokolov05,kolomeisky,mirnyarxiv,rezania,kafri}. 
\section{Brief Review of earlier models}
\label{review}
Bustamante et al. {\cite{Busta}} showed experimental evidence of the 
intersegmental transfer and hopping movements of {\it E. Coli} RNA Polymerase
(RNAP) on nonspecific DNA. They also showed
the effect of Heparin, which disrupts the RNAP-DNA nonspecific complexes. (For a
 theoretical review of this phenomenon, see {\cite{Halford1}}.)

Burdzy and Holyst {\cite{Holyst}} address an important question, namely the 
number of molecules needed to locate the target of a given size. However, the 
theoretical arguments are not supported by any simulations. Also, the arguments
 are not in terms of FPTs, which may be more relevant biologically in the 
given context. 

The effect of sequential inhomogeneity of the DNA was taken into consideration
 by Slutsky et al. {\cite{slutsky04}}. They however focussed only on a 
combination of one and three dimensional search mechanisms,
without focusing on the Intersegmental transfers. Also, they modeled the DNA as a one-dimensional strand, which is not completely realistic in the biological 
context.

The DNA was modeled as a one-dimensional strip consisting of low and high 
affinity sites by Rezania et al. {\cite{rezania}}.
They also took a two dimensional strip which, in addition to the above mentioned
sites, has zero affinity water. However, they did not investigate the role of
the bridges explicitly in their simulations.

The model developed by Oshanin et al. {\cite{oshanin}} is similar to our model, in that the search is carried out
in discrete time steps till a maximum of $N$ steps, until the immobile target is found. The survival probability
is found in terms of the leakage probability and is optimized to minimize this probability. However, the 
calculations are done for a one-dimensional substrate, which may not be 
biologically realistic.

Recently, Sheinman et al. {\cite{kafri}} studied the effect of 
intersegmental transfers on the search process. The DNA was however modeled by
connecting an ideal gas of rods (of unit persistence length) randomly to 
form a small world network. The authors reported a decrease
in the search time by using scaling arguments and numerical verification. They
also found dependence on the length of the DNA, an aspect which we do not 
address in great detail here. 

Therefore, in spite of the large attention that this problem has 
received recently, the role of all three mechanisms and, in particular, the 
role of intersegmental transfer together with the attachment/detachment 
have not been investigated thoroughly. In this paper, we study all the three 
mechanisms together, which complements some of the works which have been 
reported earlier for elucidating the relative importance of each.
\vspace{1cm}

\section{The Model}
\label{model}

A DNA can be considered to be a freely jointed chain over length scales 
much longer than its persistence length. A freely jointed chain can be 
modeled using a SAW \cite{degennes}, where the 
length of each of the steps of the SAW is typically of the order of the 
persistence length. The persistence length of DNA is roughly 100 
base-pairs (bps). Therefore, a SAW of total length $L = 100$ would 
correspond, approximately, to $10,000$ base pairs which is comparable, 
for example, to the length of a {\it bacteriophage} DNA.

Motivated by the experimental and theoretical works summarized in 
sections \ref{motivation} and \ref{review}, in this paper we explore the efficiency of 
searching the SAW by a random walker for a specific binding site on the 
SAW. We study the efficiency of various search mechanisms that the 
random walker may use in order to reach the target site. We have 
introduced a new quantitative measure of the efficiencies of the search 
mechanisms in terms of the time-scales that are relevant to this problem. 

\begin{figure}[ph]
\centerline{\psfig{file=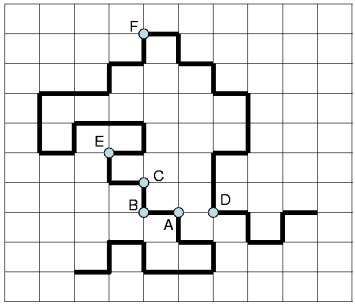,width=8.7cm}}
\vspace*{8pt}
\caption{A pictorial depiction of a Self Avoiding Walk (SAW) generated on a 
square lattice. A$\leftrightarrow$B$\leftrightarrow$C represents sliding. 
A$\leftrightarrow$D represents hopping across a {\it bridge}. A path from A to 
F which does not consist entirely of sliding along the contour and hopping 
across bridges is called jumping (It would consist of atleast one detachment 
from the SAW and its re-attachment). The probabilities associated with hopping 
from one site to the other depends on the local conformation of the SAW. 
The boundaries of the underlying lattice are far away from the 
SAW.\label{SAWsq}}
\end{figure}

For the sake of simplicity, we consider SAWs in two-dimensions, rather 
than three-dimensions. The random walker 
is represented by a particle. The particle searches the binding site 
by a combination of sliding, intersegment hopping as well as detachments, 
two-dimensional diffusion followed by, possibly, re-attachments (see 
Fig.\ref{SAWsq}). {\it Sliding} motion of the particle is captured by 
its one-dimensional RW where its position at the successive time steps 
are nearest-neighbours along the contour of the SAW. In contrast, an 
{\it inter-segment hopping} of the particle takes place across a 
``bridge'' that connects two sites both of which are located on the 
SAW and are nearest-neighbours on the square lattice but are not nearest 
neighbours along the contour of the SAW. Finally, upon {\it detachment} 
from the SAW, a particle executes an unbiased RW on the square lattice 
and, during this process, may {\it re-attach} with the SAW if it hops 
onto a site occupied by the SAW. 

In our model we generate SAW configurations, each of length $L = 101$, on a 
square lattice (Fig.{\ref{SAWsq}}) using a combination of {\it reptation} and the {\it kink 
jump} algorithms \cite{krembind}. Averaging over the configurations thus generated, we have 
verified that the {\it mean-square end-to-end Euclidean distance of the SAWs} satisfy the 
well known relation $<r_L^2> {\propto} L^{3/2}$. When the random walker was 
constrained to move only along the SAW, it performed, effectively, one-dimensional diffusion. 
We can determine the value of the effective diffusion constant 
$D$, where $D=<R_{t}^{2}>/(2t)$, $<R_{t}^{2}>$ being the mean square displacement {\it along the contour} 
of the SAW. We have also verified that the mean-square {\it Euclidean displacement} of the 
random walker, on the SAW, follows $<R_E^2(t)> \propto t^{3/4}$, even when hopping across the bridges are 
allowed. This is in agreement with the results reported earlier 
\cite{chowdhury85a,chowdhury85b}. 
\section{Results and Discussion}
\label{result}

We parametrize the positions along the contour of the SAW by 
the symbol $s$; $s = 1$ and $s = L$ correspond to the two end points on the SAW. We designate the two end points, 
i.e., $s = 1$ and $s = L$ as the specific binding sites for the particle. 
On each SAW of length $L$, we release a particle at the mid-point of the SAW
(i.e., at $s = (L+1)/2$) and allow it to execute a RW for a total of $N$ discrete
time steps. If the particle is unable to reach either of the target sites 
(i.e., $s=1$ or $s=L$), then
the search by that particle is aborted and the search by another particle
 starts again. $N$ is $5000$ and 
$L$ is $101$ in all our simulations. In three different sets 
of computer experiments we implemented three different types of RWs of the particle.\\ 
(i) Mechanism I (M I): The particle is allowed to perform random walk only along the contour of the SAW.\\
(ii) Mechanism II (M II): Hopping across the bridges is allowed, in addition to the process included in mechanism I \cite{allowed}.\\ 
(iii) Mechanism III (M III): Attachment and detachment of the particle are also allowed, in addition to the processes 
included in mechanism II \cite{allowed}. 

For the random walkers, we impose absorbing boundary 
conditions at $s = 1$ and $s = L$, i.e. a succesful search process is terminated
once the walkers reach the target site for the first time. Under these boundary
conditions, the time taken by a random walker to reach one of the two 
boundaries (i.e., $s = 1$ or $s = L$) is identified as the corresponding FPT.\\

\subsection{Distributions of First Passage Times(FPTs)}

\begin{figure} 
  \begin{center}
    \begin{tabular}{cc}
      \resizebox{80mm}{!}{\includegraphics{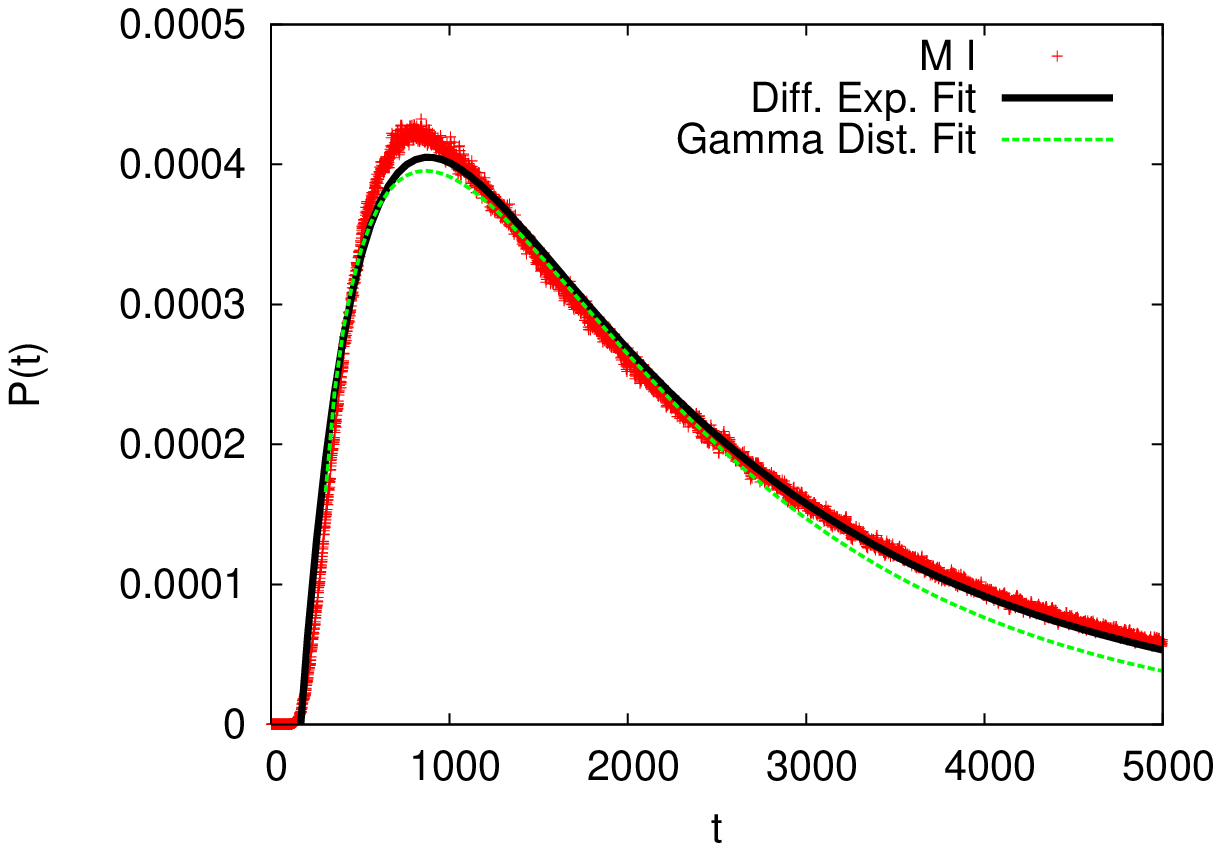}} \\
      \resizebox{80mm}{!}{\includegraphics{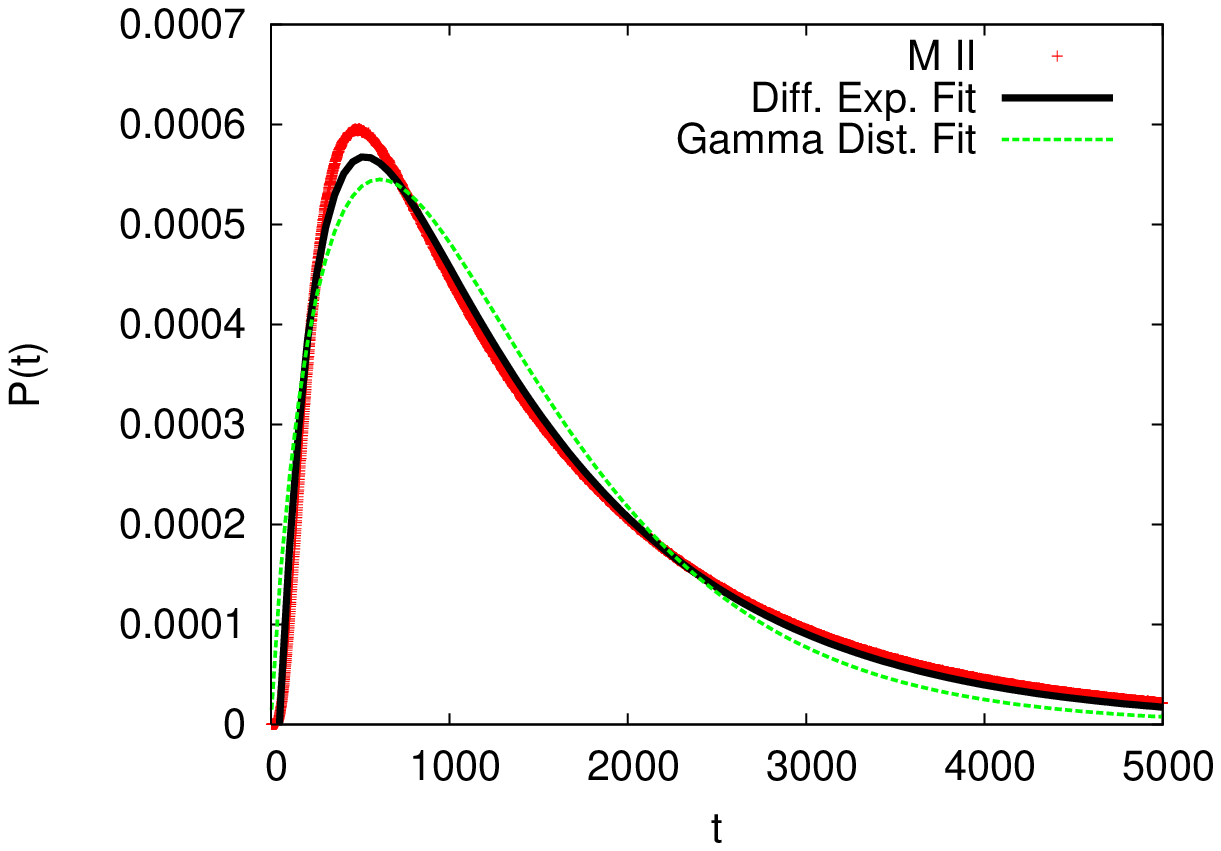}} \\
      \resizebox{80mm}{!}{\includegraphics{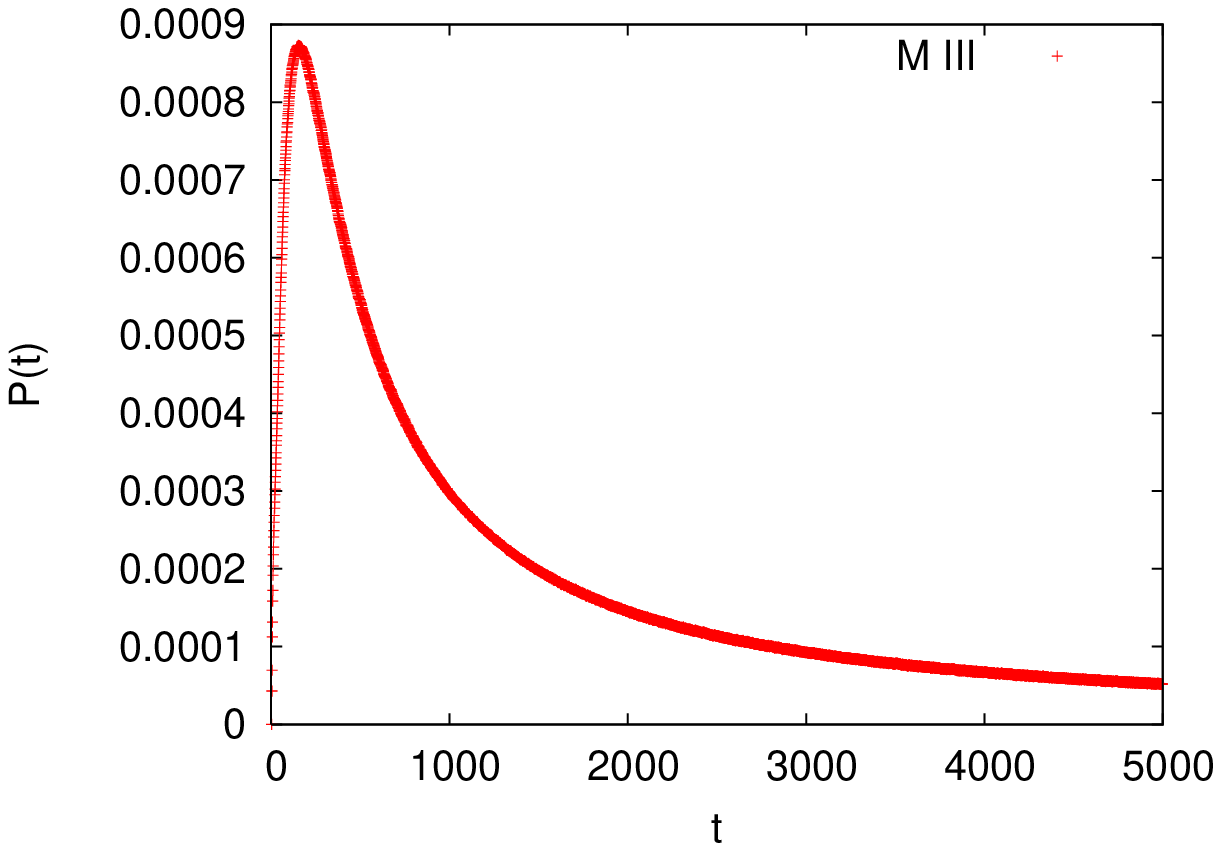}} \\
    \end{tabular}
    \caption{The distribution of the FPTs for 
(a) Mechanism I, (b) Mechanism II, and (c) Mechanism III. (See Appendix {\rm I}
 for fit parameters). }
    \label{comparemech}
  \end{center}
\end{figure}

The distribution $P(t)$ of the FPTs for the three mechanisms are plotted in Fig.\ref{comparemech}. 
Since all three mechanisms are based on diffusive search, the qualitative shape of the curve $P(t)$ is the 
same in all the three cases. But, comparing the most probable time for three mechanisms, we conclude that 
the mechanism II is more efficient than mechanism I whereas mechanism III is the most efficient of all. This observation  
strongly suggests that the search for DNA-binding sites by proteins would be more efficient if, in addition to 
sliding, both inter-segment hopping and detachment/re-attachment are also allowed. 

\subsection{Relative importance of detachments/re-attachments}

\begin{figure}[ph]
\centerline{\psfig{file=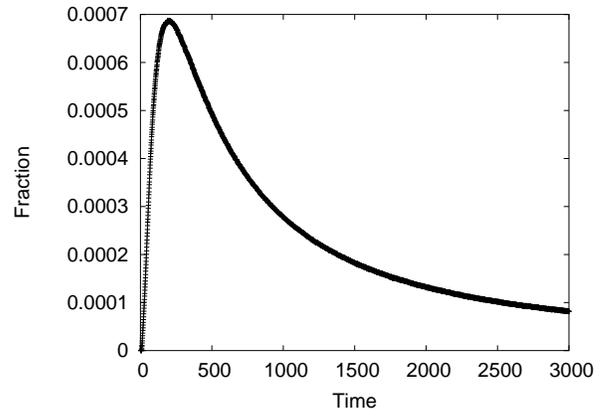,width=7.7cm}}
\vspace*{8pt}
\caption{The fraction of the time steps during which the particle remains unattached from the SAW before reaching 
the target binding site in $t$ steps.\label{fracunat}}
\end{figure}

In order to compare the relative importance of detachment/re-attachment compared to sliding and inter-segment 
hopping, we have computed the fraction of the time steps the particle spends unattached with the SAW in each 
successful search process. Corresponding to every search time, $t$, we compute
the fraction of the search time that the particle spends unattached from the 
SAW. We plot this fraction as a function of the search time in Fig.\ref{fracunat}  
. Note that the peak in Fig.{\ref{fracunat}} occurs at $t=191$. Interestingly,
this value is close to the most probable FPT in Fig.{\ref{comparemech}},
corresponding to Mechanism III, namely $t=153$. 
Thus, the target site is reached in the shortest possible time if the particle uses
mechanism III,
in which the searching particle spends a fraction of the search time  
outside the SAW.

\begin{figure}
\begin{center}
\psfig{figure=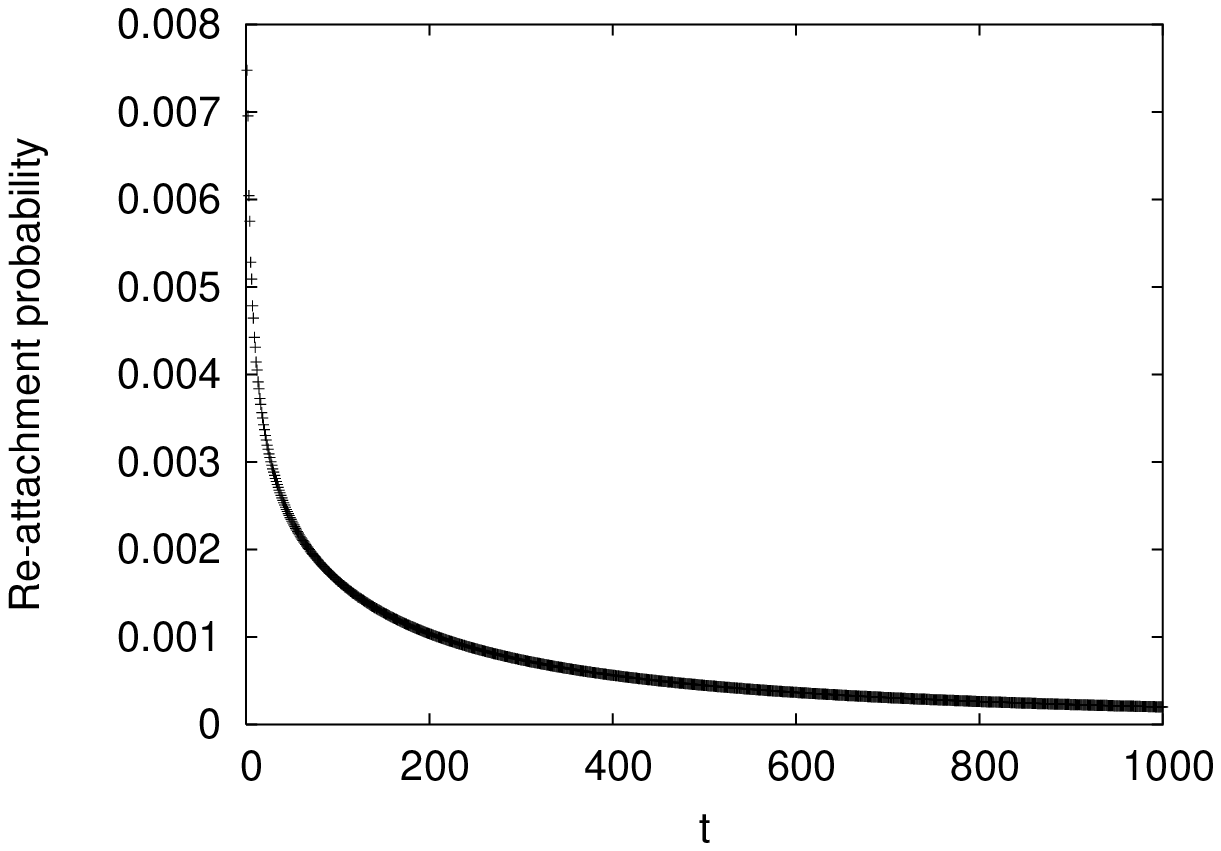,width=90mm}
\vspace*{-100mm}
\end{center}
\vspace*{45mm}
\hspace*{35mm}
\psfig{figure=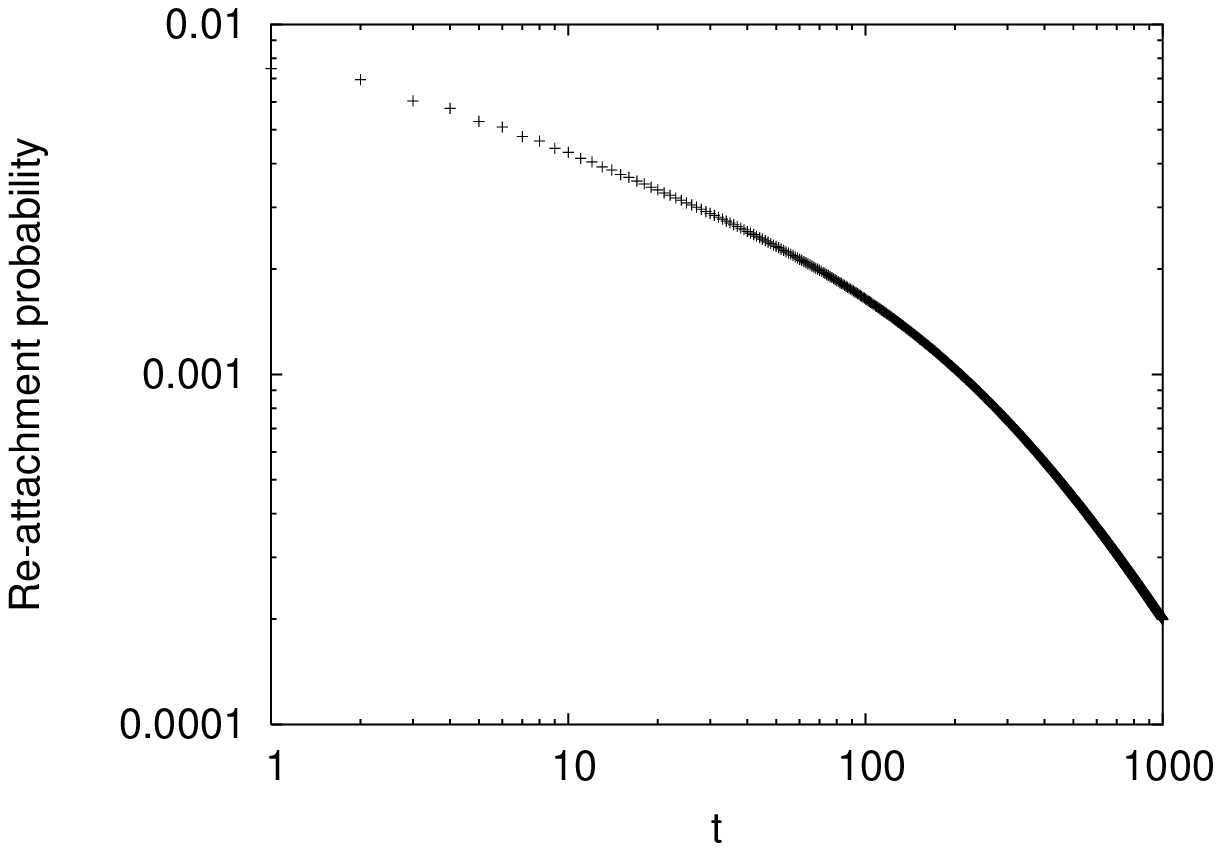,width=45mm}
\vspace*{30mm}
\caption{The re-attachment probability in Mechanism III. The inset shows the same data on a log scale}
\label{reattach}
\end{figure}

We have also computed the probability of re-attachment of a particle after $t$ time steps, following its 
detachment from the SAW; this probability distribution is shown in 
Fig.\ref{reattach}. The log-log plot in the inset indicates the 
possibility of an initial {\it power law} regime, which is most likely $\sim t^{-1/2}$, crossing over to 
another power law regime at long times, which was found to be $\sim t^{-3/2}$.  

\subsection{Mechanism I versus Mechanism II}

In this subsection, we consider a {\it modified version} of Mechanism II (MM II) 
which reduces to the mechanism I in a special limit. 
In this modified version, we compute the effect of forced hopping across 
the bridges, with a given probability.  We define a quantity $R$ as follows,
\begin{eqnarray}
R={\dfrac{p_{bridge}}{p_{contour}}} \\
p_{bridge}+p_{contour}=1
\label{R}
\end{eqnarray}

where $p_{bridge}$ is the probability of hopping across the bridge and 
$p_{contour}$ is the probability of diffusing along contour. In the 
limit $p_{bridge} = 0$ (i.e., $R = 0$), this modified version reduces to 
mechanism I.

\begin{figure}
  \begin{center}
    \begin{tabular}{cc}
      \resizebox{80mm}{!}{\includegraphics{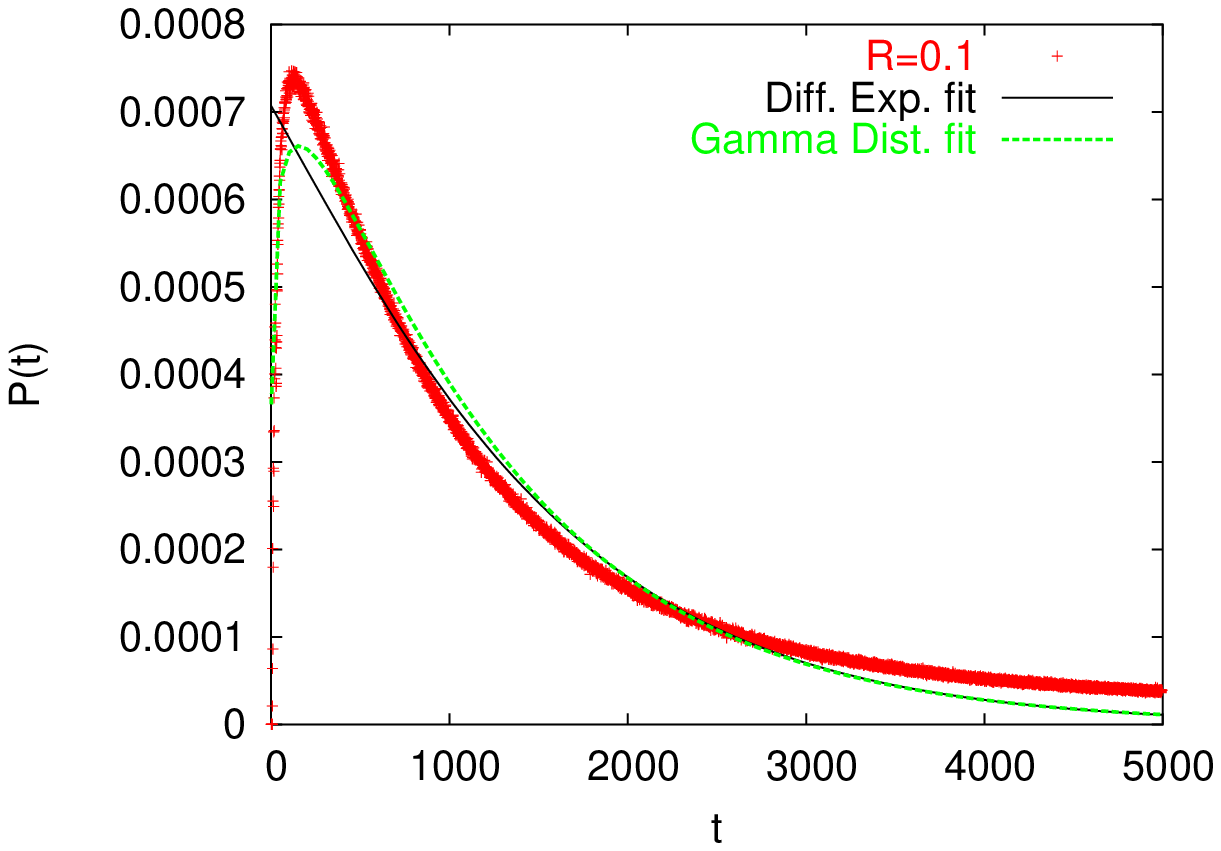}} \\
      \resizebox{80mm}{!}{\includegraphics{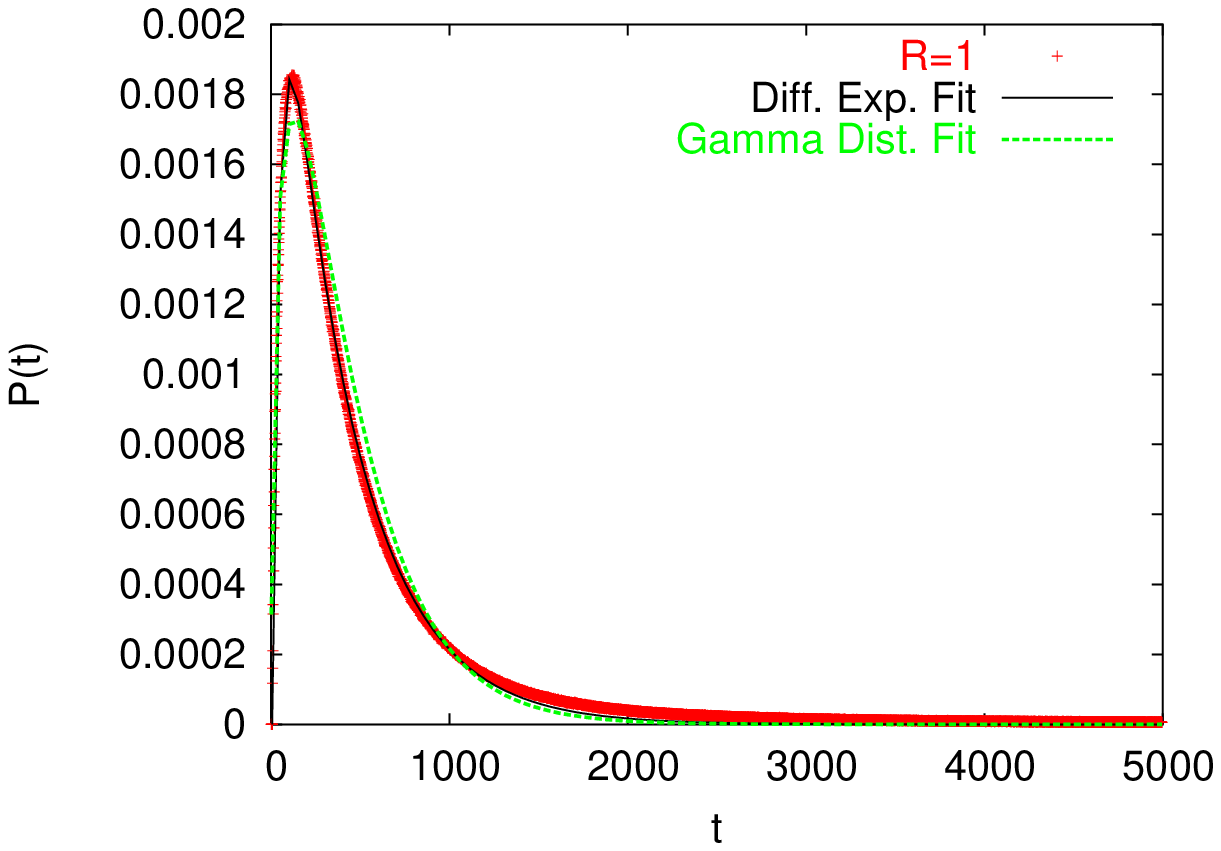}} \\
      \resizebox{80mm}{!}{\includegraphics{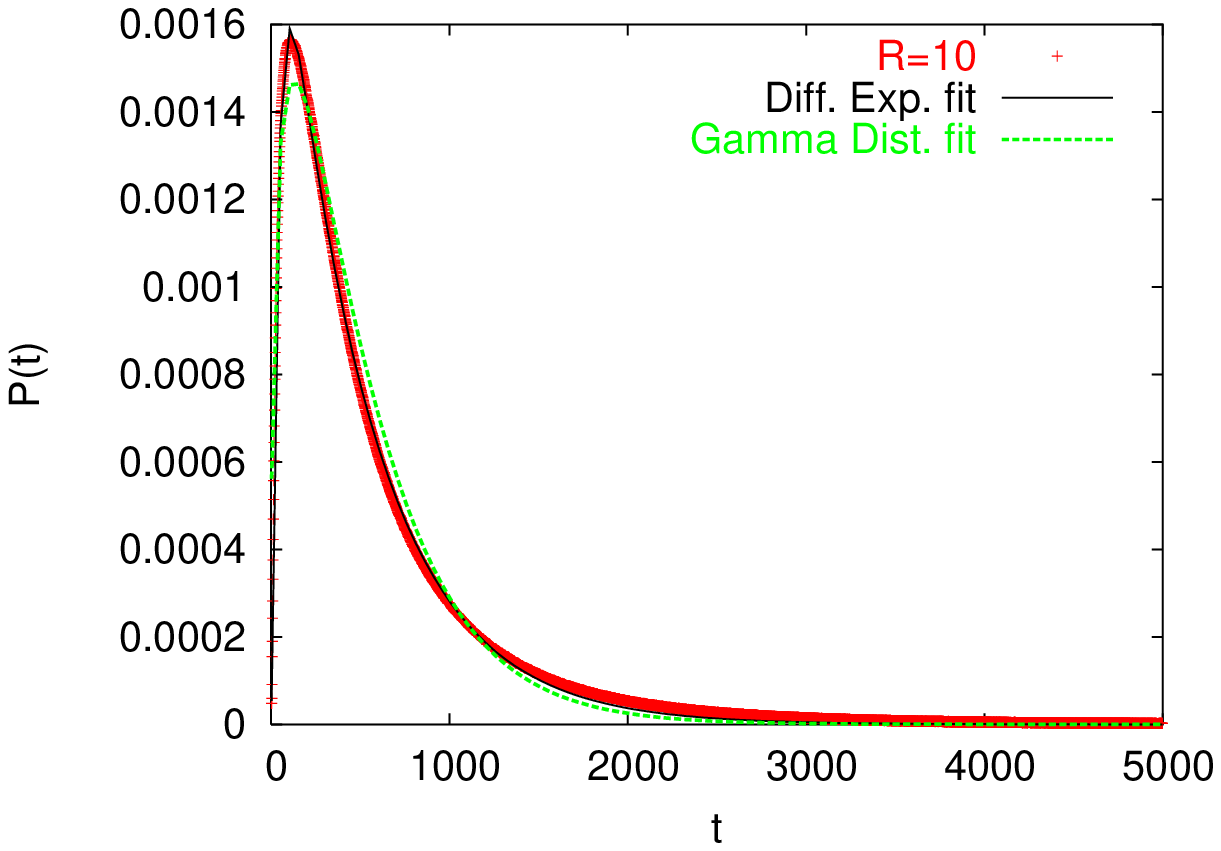}} \\
       \resizebox{80mm}{!}{\includegraphics{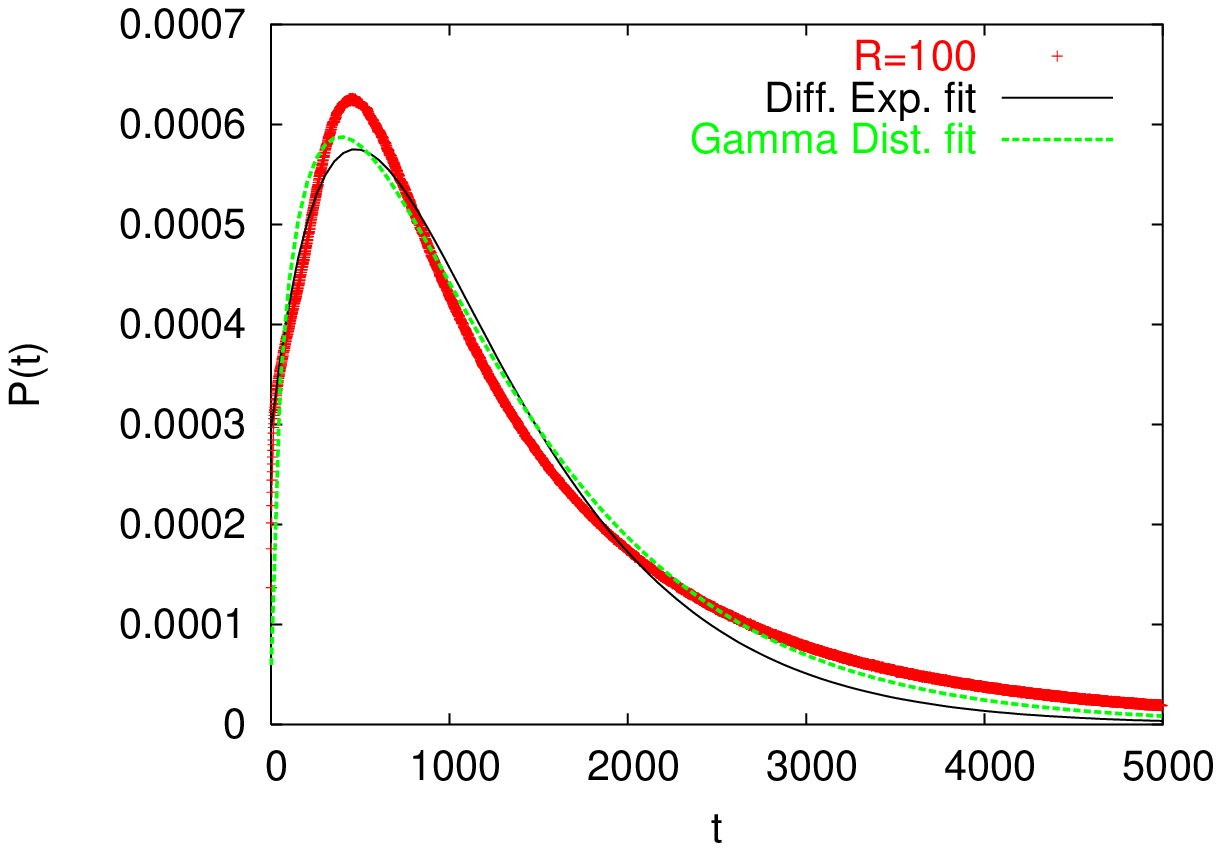}} \\
    \end{tabular}
    \caption{The distribution of the FPTs for the modified Mechanism II for 
(a) R=0.1, (b) R=1, (c) R=10, and (d) R=100. To each curve, we fit a Gamma 
Distribution and a Difference of Exponentials (see Appendix {\rm I}). }
    \label{Rplots}
  \end{center}
\end{figure}

In Fig.({\ref{Rplots}}), we plot the distribution $P(t)$ of the FPTs for 
four different values of $R$.   

A higher value of $R$ indicates a higher probability of hopping across a 
bridge. This gives rise to a higher probability of reaching the ends in 
roughly the same amount of time. Therefore, if the protein has some bio-chemical
means of hopping across such bridges preferentially, then it can bind to the
specific binding site in a more efficient manner.

However, as we see from Fig.{\ref{Rplots}}, for an extremely high value of 
$R$, the walker tends to get trapped in the bridge and hence takes a longer 
time to reach the ends. For example, when $R=100$, $p_{bridge}\approx 0.99$.
For this value of $p_{bridge}$, the moment the walker encounters a bridge, it 
would tend to get trapped in a bridge between two sites (for example, the 
bridge connecting ``A" and ``D" in Fig. {\ref{SAWsq}}). 

\subsection{Quantitative estimates of efficiencies of search-times}

We are now in a position to compare the values of the most probable time,
$\tau_{mp}$, and the MFPT ($\tau_{avg}$) for the distribution of the FPTs of 
all the mechanisms that we have investigated till now. Let $\tau_{1D}$ be the 
most probable/ MFPT for successful search using
Mechanism I while $\tau$ be the corresponding most probable/ MFPT
for the specific mechanism under consideration.

We define 
\begin{equation}
{\eta}=|1-\frac{\tau}{\tau_{1D}}|,
\label{eta}
\end{equation}
which we use as a quantitative 
measure of the efficiency of the process, relative to purely one-dimensional
diffusion. 
The data are summarised in the table below.

\begin{table}[ht]
{\begin{tabular}{@{}ccccc@{}} \toprule
{\bf Mechanism} & {\bf Most Probable} & {\bf $\eta_{mp}$}&{\bf Mean Search}&{\bf $\eta_{avg}$}\\ 
&  {\bf Search Time ($\tau_{mp}$)} & &{\bf Time ($\tau_{avg}$)} &\\ \colrule
Mechanism I&841&0&1931.6&0\\
Mechanism II&494&0.41&1419.8&0.27\\
Mechanism III&153&0.82&1346.5&0.30\\
MM II (R=0.1)&115&0.86&1243.9&0.36\\
MM II (R=1)&117&0.86&553.7&0.71\\
MM II (R=10)&112&0.87&598.3&0.69\\
MM II (R=100)&457&0.46&1243.9&0.36\\ \botrule
\label{tablemp}
\end{tabular}} 
\end{table}
We conclude that among the possible mechanisms
considered in this paper, the modified Mechanism II with 
$R=10$ turns out to be the most efficient search process, as far as $\eta_{mp}$
is concerned. However, in terms of $\eta_{avg}$, $R=1$ would be the most 
efficient search mechanism. Therefore, we conjecture that if both $\eta_{mp}$
and $\eta_{avg}$ play equally important roles in determining the efficiency of 
a given mechanism, then the most efficient search mechanism would correspond to
the range $1\leq R\leq 10$. 

\begin{figure}
  \begin{center}
    \begin{tabular}{cc}
      \resizebox{90mm}{!}{\includegraphics{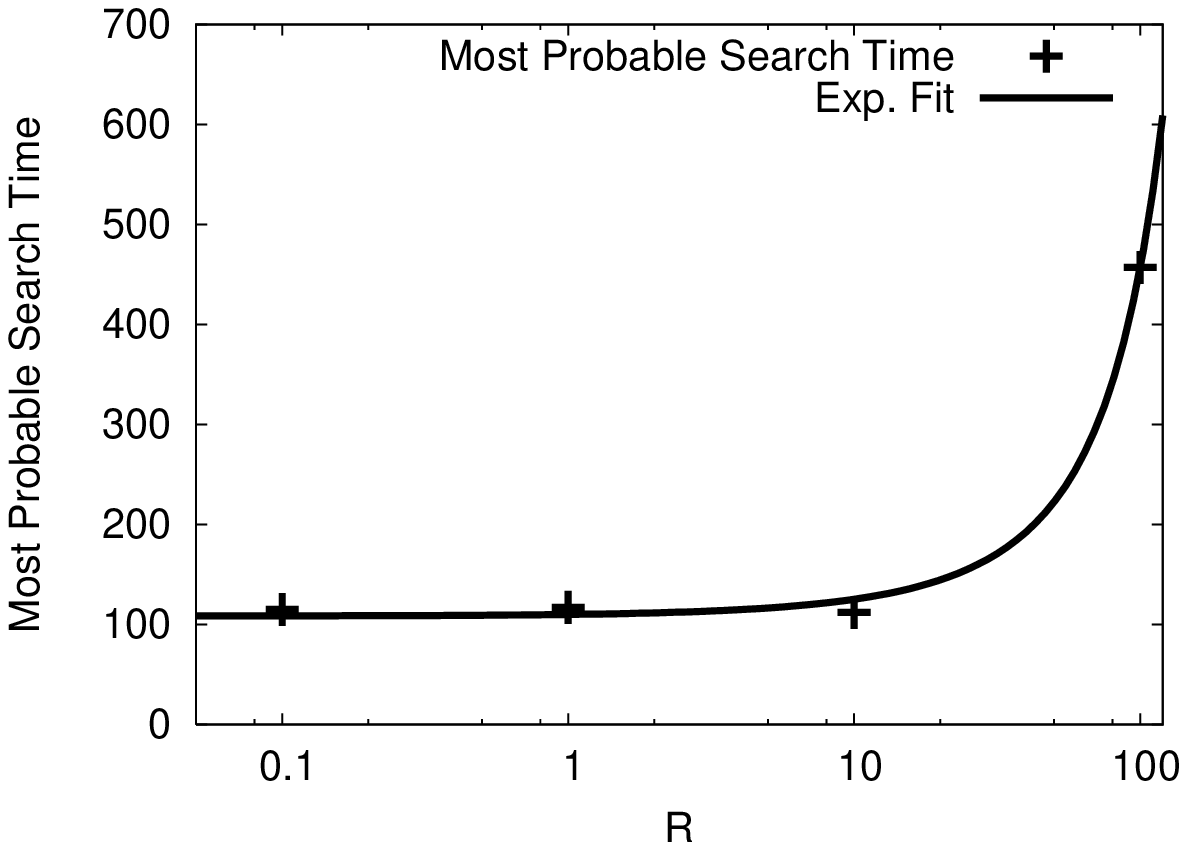}} \\
      \resizebox{90mm}{!}{\includegraphics{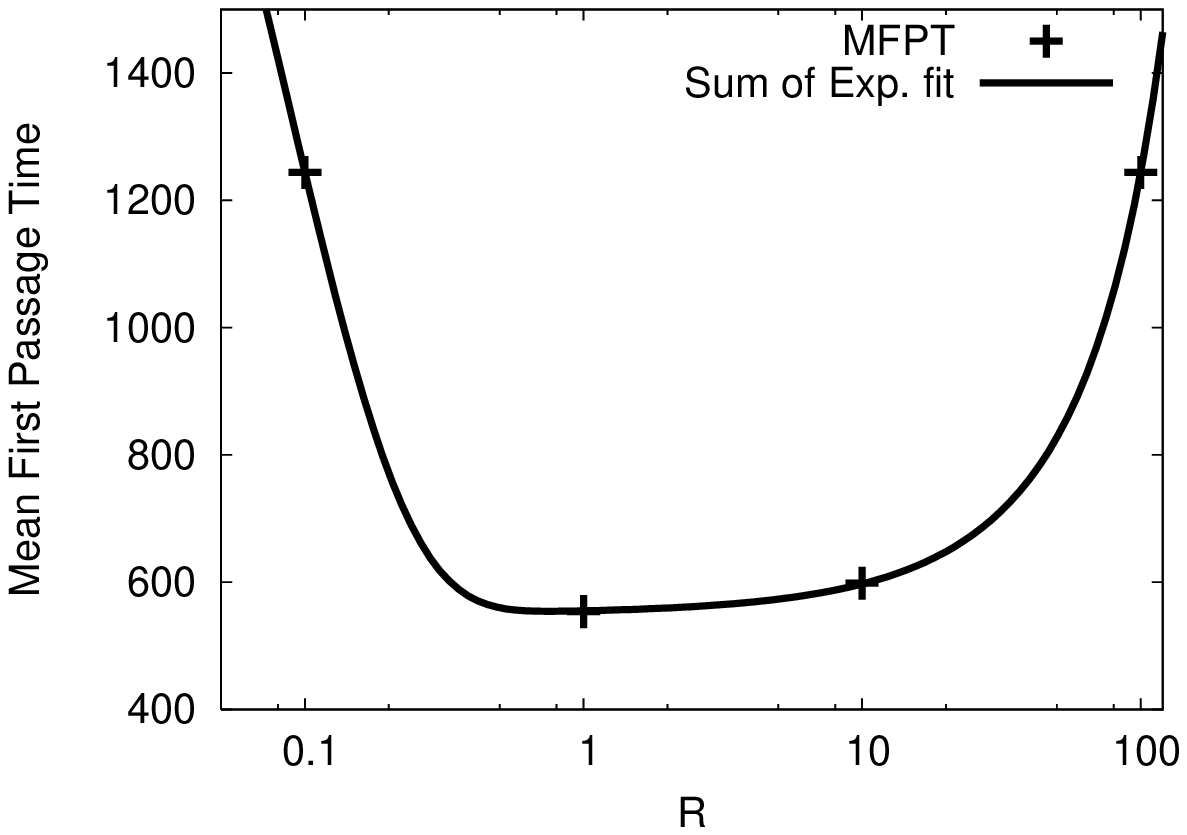}} \\
    \end{tabular}
    \caption{(a.) The $\tau_{mp}$ as a function of $R$. (b.) The $\tau_{avg}$ as a function of $R$. (See Appendix {\rm I} for fit parameters) }
    \label{taumpavg}
  \end{center}
\end{figure}


\begin{table}[ht]
{\begin{tabular}{@{}ccc@{}} \toprule
{\bf Mechanism}&$a\approx$&$b\approx$\\ \colrule
M I&1.483&0.001\\
($k\approx 266.725$)&&\\
M II&1.836&0.001\\
MM II (R=0.1) \hphantom{00} &\hphantom{00} 1.146 \hphantom{00} &\hphantom{00} 0.001\\
MM II (R=1) \hphantom{00} &\hphantom{00} 1.442 \hphantom{00} &\hphantom{00} 0.003\\
MM II (R=10) \hphantom{00} &\hphantom{00} 1.346 \hphantom{00} & \hphantom{00} 0.003\\
MM II (R=100) \hphantom{00} &\hphantom{00} 1.461 \hphantom{00} & \hphantom{00} 0.001\\ \botrule
\end{tabular} \label{tablab}}
\end{table}
\begin{table}[ht]
{\begin{tabular}{@{}ccccc@{}} \toprule

{\bf Mechanism}&$c\approx$&$d\approx$&$f\approx$&$h\approx$\\ \colrule
M I&\hphantom{00}-0.001&\hphantom{00}0.003&\hphantom{00}-0.001&\hphantom{00}0.001\\
M II&\hphantom{00}0.001&\hphantom{00}0.001&\hphantom{00}0.001&\hphantom{00}0.004\\
MM II (R=0.1)&\hphantom{00}0.100&\hphantom{00}0.001&\hphantom{00}0.099&\hphantom{00}0.001\\
MM II (R=1)&\hphantom{00}-0.003&\hphantom{00}0.023&\hphantom{00}-0.003&\hphantom{00}0.003\\
MM II (R=10)&\hphantom{00}-0.002&\hphantom{00}0.026&\hphantom{00}-0.002&\hphantom{00}0.002\\
MM II (R=100)&\hphantom{00}0.996&\hphantom{00}0.002&\hphantom{00}0.996&\hphantom{00}0.002\\ \botrule
\end{tabular} \label{tablcdfh}}
\end{table}

We also observe another interesting feature in MM II. We note that $P_{max}=
P(\tau_{mp})$ is largest for $R=1$ and then for $R=10$. This implies that not only are
they the most efficient of all the mechanisms considered, but that they also have the 
highest ``success-rate" of reaching the taget sites. Therefore, we see that MM II with forced
hopping across the bridges leads to the most efficient and successful search process. 
Whether all proteins with multiple DNA-binding sites actually make use of this mechanism to
reach their target sites is something that needs to be tested experimentally under controlled 
conditions in the near future. 

In Fig. {\ref{taumpavg}}, we plot the most probable search time and the 
mean first passage time as functions of $R$. We do not show the point $R=0$ 
(in the limit $R\to 0$, we recover $\tau_{mp}=841$ and $\tau_{avg}=1931.6$)
on the log-scale. The only quantitative difference between the two is that the
turning point in the curve for the MFPT lies in the range $0.1\leq R\leq 1$,
whereas, the turning point in the curve for the $\tau_{mp}$ lies in the range
$0\leq R\leq 0.1$. 

\subsection{Multiple Walkers and Immovable Barriers}
We have also investigated the search of the same binding sites simultaneously by $N$ ($>1$) 
interacting particles which are initially distributed randomly along the SAW. The positions 
of the particles are updated in parallel subject to the constraint that none of the lattice 
sites is occupied by more than one walker at a time. As is suggested by our intuition, the 
$<R_t^2>$ decreases with an increasing number of random walkers. In case of pure sliding, 
the interaction between the particles, effectively, constrains each one to a shorter region 
on the SAW. Consequently, $<R_t^2>$ decreases with increasing $N$. However, in the presence 
of {\it Bridges}, there arise some situations in which a particle can {\it bypass} the 
other particles on its way by hopping across the bridges and, thereby, increasing $<R_t^2>$. Effects
of mutual hindrance is further weakened by detachments/re-attachment processes.

We also considered the situation when there are immovable barriers placed 
randomly along the SAW. This could mimic the effect of various obstacles that
are present {\it in-vivo} in the crowded environment of the cell. 
The mechanism III is the most efficient search
process in the presence of these barriers.   

\section{Summary and Conclusion}
In this paper, we have suggested a biologically motivated extension of random walk on 
self-avoiding walks. The results of this investigation provide insight into the 
relative importance of different mechanisms of search for specific binding on DNA by 
DNA-binding proteins. We studied the effect of preferential bias to hop across the 
bridges in the intersegmental transfer and found that for $1\leq R\leq 10$,
the mechanism II turns out to be most efficient. Whether this is the mechanism that
proteins actually use in order to find the target sites can be verified only by doing
controlled experiments.  

We also suggest experiments that can be performed to test
the efficiency of the various search processes. The value of $\tau_{1D}$ can
be taken as an input from standard known results. The value of $\tau_{mp}$
and $\tau_{avg}$ can be measured using {\it Fluoroscence Spectroscopy}. 
The experimentally obtained $\eta$ can then be compared with the above mentioned results 
,obtained using simulations to throw light on the possible mechanism that the 
protein uses to search for its target site.
\section*{Acknowledgments}
The idea of this work originated while visiting
the Max-Planck Institute for the Physics of Complex Systems, Dresden during 
summer '07. 

I thank the visitors programme of MPI-PKS for the hospitality in Dresden. 
I thank S.W. Grill, E.A. Galburt, A.B. Kolomeisky, B.K. Chakrabarti, Abhishek Dhar, 
Abhishek Chaudhuri, Aditya Sood and Ashok Garai for fruitful
discussions and comments. I also thank Debashish Chowdhury, R. Metzler and 
J. Klafter for drawing my attention to some relevant earlier works. 

\appendix
                                                                                
\section{}
In this appendix, we analyze the FPT distributions obtained for all the 
mechanisms, quantitatively. We know that the {\it Gamma} distribution (GD) is
one of the most appropriate forms for modelling waiting time distributions
and other similar phenomena.
We fit all our FPT distributions (apart from M I) using a two parameter GD:
${\cal{F}}(t)=b^{a} t^{a-1} e^{-bt}/{\Gamma}(a)$, where $\Gamma (a)$ is the 
gamma-function of $a$, while  $a$ and $b$ are 
parameters to be fitted using least squares regression. For Mechanism I, we
fit the FPT distribution to ${\cal{\tilde{F}}}(t)=b^{a} (t-k)^{a-1} e^{-b(t-k)}/{\Gamma}(a)$, where $k$ is also a parameter to be fitted using least squares 
regression.

We observe that the data for the FPT distribution fits equally well
to the difference of two exponentials. We fit the distributions to a 
four parameter function as follows : ${\cal{F}}(t)=c e^{-d t} - f e^{-h t}$, 
where $c$, $d$, $f$ and $h$ are the parameters to be fitted using least 
squares regression. Both the GD and the difference of exponential fits 
to the FPT distribution of Mechanism III were poor, and hence not shown in the
figure.

We have listed the fit parameters in the tables in Section{\ref{tablab}}.

In Fig.{\ref{taumpavg}}, we plot the $\tau_{mp}$ and $\tau_{avg}$ as functions
of $R$. We find $\tau_{mp}=a e^{b R}$, where $a\approx 108.343$ and 
$b\approx 0.014$. On the other hand, $\tau_{avg}=c e^{d R} + f e^{h R}$,
where $c\approx 550.342$, $d\approx 0.008$, $f\approx 2180.460$ and 
$h\approx -11.462$.


%


\begin{thebibliography}{0}
\bibitem{allowed} In Mechanism II, for example, the probability of jumping from
site A to site D (see Fig. \ref{SAWsq}) is $\frac{1}{3}$, whereas the probability of jumping from D to 
A is $\frac{1}{4}$. For mechanism III, the probability of detachment varies 
from site to site. The probability of jumping from any site to the $4$ 
neighboring sites is $\frac{1}{4}$ each.
Therefore, for sites �A,�B�and C it is $\frac{1}{4}$, whereas for site F, it is
$\frac{1}{2}$. Also, for site D, it is $0$.
\end{thebibliography}
\end{document}